\documentclass{rnaastex}

\begin{document}

\title{High-Spatial-Resolution K-Band Imaging of Select K2 Campaign Fields}

%% Note that the corresponding author command and emails has to come
%% before everything else. Also place all the emails in the \email
%% command instead of using multiple \email calls.
\correspondingauthor{Knicole D. Col\'on}
\email{knicole.colon@nasa.gov}

\author{Knicole D. Col\'on}
\affiliation{NASA Goddard Space Flight Center, Greenbelt, MD 20771, USA}

\author{Steve B. Howell}
\affiliation{NASA Ames Research Center, Moffett Field, CA 94035, USA}

\author{David R. Ciardi}
\affiliation{NASA Exoplanet Science Institute/Caltech, Pasadena, CA
91125, USA}

\author{Thomas Barclay}
\affiliation{NASA Goddard Space Flight Center, Greenbelt, MD 20771, USA}

%% Note that RNAAS manuscripts DO NOT have abstracts.

%% Start the main body of the article. If no sections in the 
%% research note leave the \section call blank to make the title.
\section{} 

\begin{figure}[h!]
\begin{center}
\includegraphics[scale=0.39,angle=0]{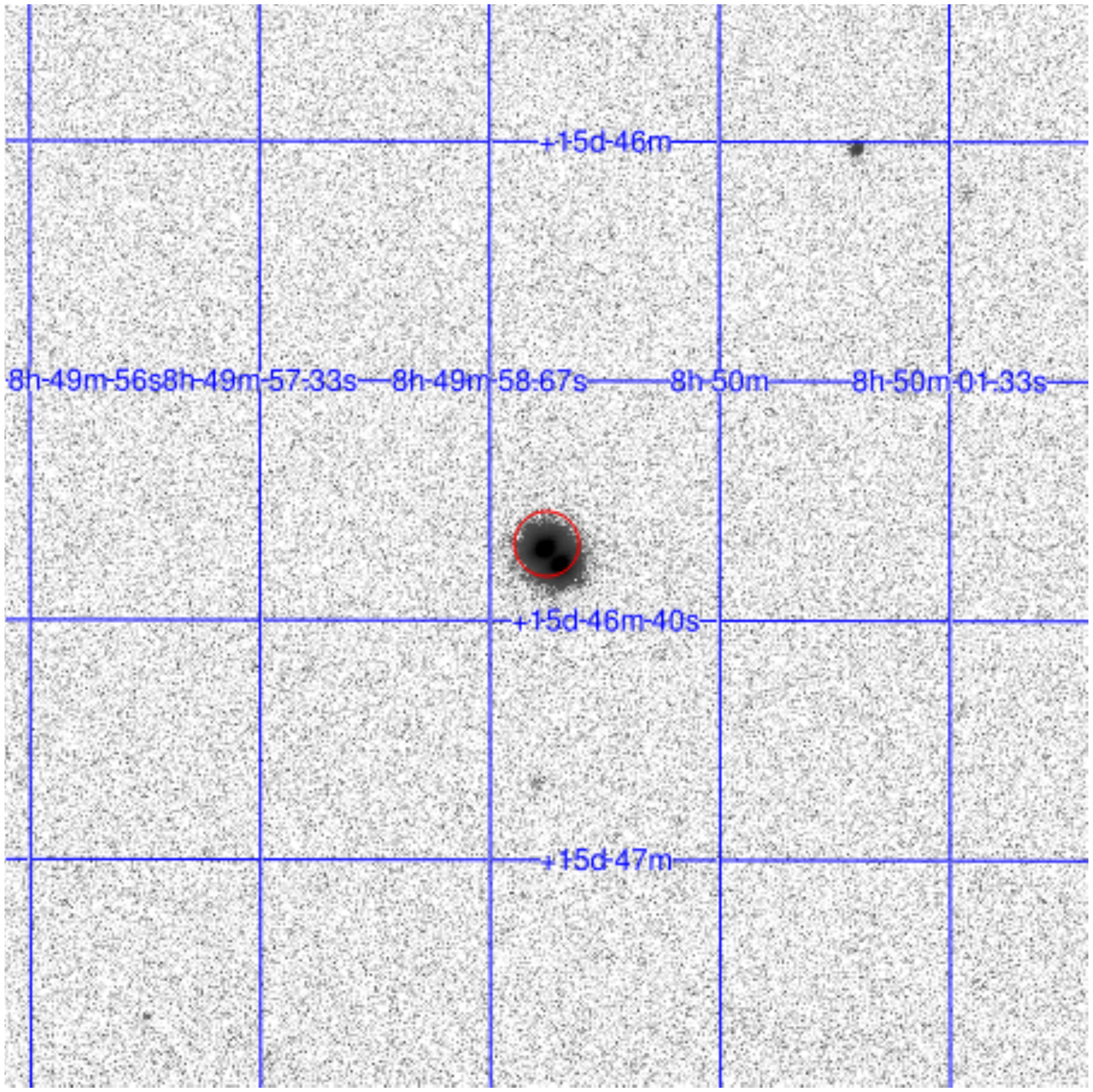}
\includegraphics[scale=0.39,angle=0]{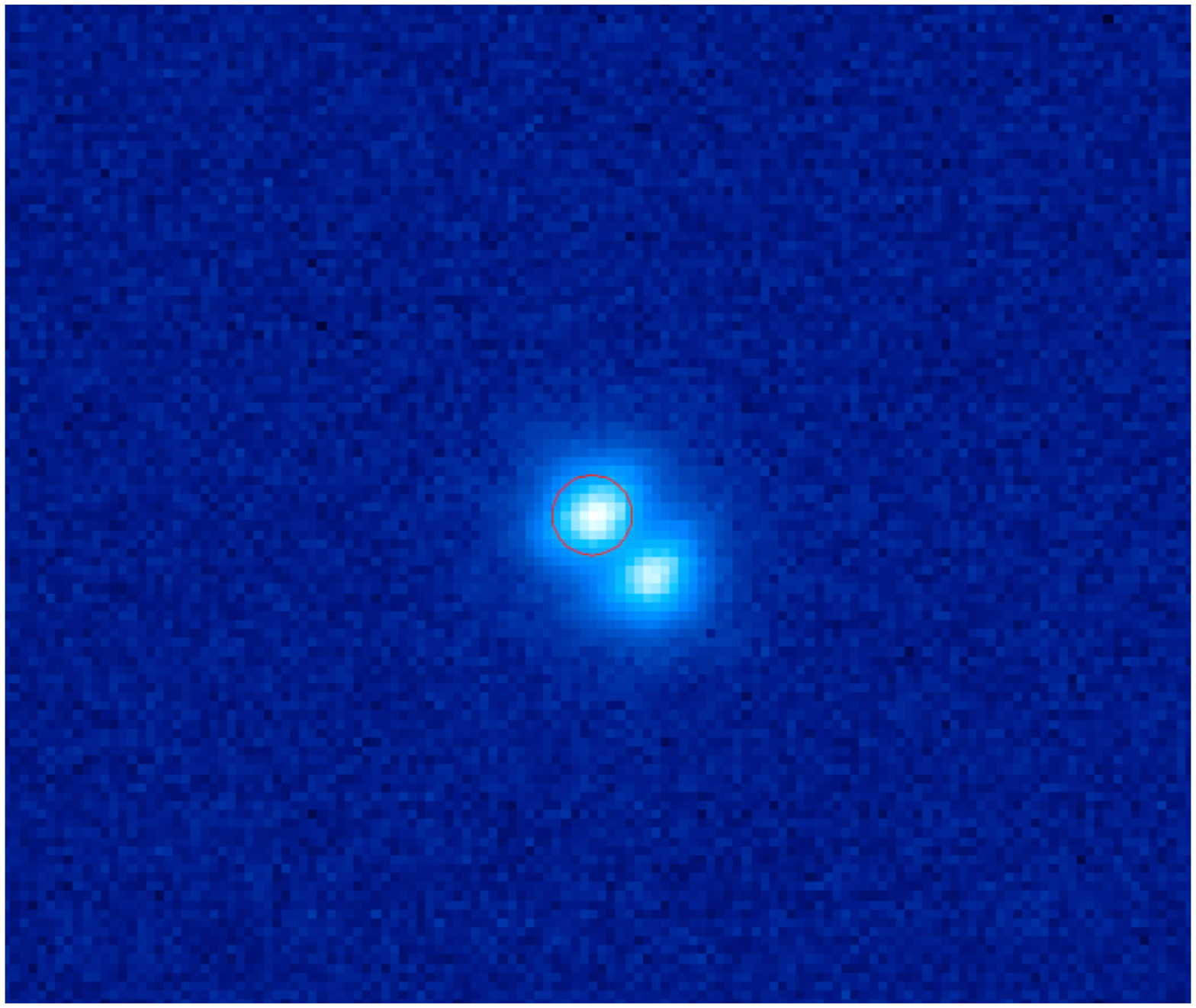}
\caption{WFCAM K-band images of EPIC 211694226, which was observed by K2 in C5 and will be observed again in C16. The left-hand image is the same format as the PNG cutouts available on the ExoFOP website. The right-hand image is a zoomed-in cutout created from the FITS file that is also available on the ExoFOP website. This target was identified as hosting a $\sim$1 $R_{\oplus}$ planet candidate in \citet{Dressing2017}; however, the authors noted it has a nearby neighbor, which the WFCAM image clearly reveals to be nearly equal in brightness as the target. \label{fig:1}}
\end{center}
\end{figure}

NASA's K2 mission began observing fields along the ecliptic plane in 2014 \citep{Howell2014}. Each observing campaign lasts $\sim$80 days, during which high-precision optical photometry of select astrophysical targets is collected by the Kepler spacecraft. Due to the 4 arcsec pixel scale of the Kepler photometer, significant blending between the observed targets can occur (especially in dense fields close to the Galactic plane). We therefore undertook a program to use the Wide Field Camera (WFCAM) on the 3.8 m United Kingdom InfraRed Telescope (UKIRT) \citep{Casali2007} to collect high-spatial-resolution near-infrared images of targets in select K2 campaign fields, which we report here. These 0.4 arcsec resolution K-band images offer the opportunity to perform a variety of science, including vetting exoplanet candidates by identifying nearby stars blended with the target star (Figure \ref{fig:1}) and estimating the size, color, and type of galaxies observed by K2.

The UKIRT/WFCAM observing program was originally designed to capture a majority of the targets observed in K2 Campaigns (C) 4 and 5.\footnote{The locations of K2 campaigns on sky can be viewed here: https://keplerscience.arc.nasa.gov/k2-fields.html} However, since there is some overlap in C4 and C13 fields, we naturally acquired WFCAM images of some C13 targets. In addition, C16 and C18 (which have not taken place yet to date) have been specifically designed to overlap significantly with the C5 field. Therefore, we also have WFCAM images for targets in C16 and C18. {\it In total, we have WFCAM images for 14,722 of 15,781 C4 targets, 18,437 of 26,527 C5 targets, 528 of 21,367 C13 targets, and 13,844 of 29,413 C16 targets.} The official target list for C18 has not yet been released so we do not know how many of those targets have WFCAM images. %In addition, we note that some targets were observed multiple times, but only the unique number of targets with WFCAM images is listed here.

The WFCAM observations were conducted between December 2015$-$January 2016 as part of proposal U/15B/NA04 (PI: Howell). WFCAM consists of 4 Rockwell Hawaii-II (HgCdTe 2048$\times$2048) arrays, where each of the individual arrays covers 13.65 $\times$ 13.65 arcmin. For all observations, we used the K filter \citep{Hodgkin2009} and survey observation mode, with a 2-point 3.2 arcsec jitter and 5 s, 1 co-add exposures with a 2$\times$2 small microstepping sequence. This microstepping sequence involved taking 4 integrations, offset by 1.42 arcsec along each axis, in order to critically sample the WFCAM point spread function.  The total integration time for each image is 40 s, providing completeness to at least K$\sim$17.5.

The raw WFCAM data was processed by the Cambridge Astronomical Survey Unit (CASU). A description of the processing steps can be found in \citet{Irwin2004}. The processed data was then transferred to the WFCAM Science Archive (WSA) and archived \citep{Hambly2008}. A total of 1453 stacked pawprint FIT images were collected in our program. We note all four arrays on WFCAM are contained within a single stacked pawprint image. Confidence maps and object catalogues are also generated by the CASU pipeline. The full data set can be accessed at the WSA\footnote{http://wsa.roe.ac.uk/} or by contacting the corresponding author. %The total size of all the data files combined is $\sim$154 GB.

For ease of viewing and analysis, we extracted 1.5$\times$1.5 arcsec cutouts from the WFCAM images around the targets for which K2 pixels were downlinked. These cutouts have been made publicly available at the ExoFOP-K2 website\footnote{https://exofop.ipac.caltech.edu/k2/} as both PNG and FITS files. An example is shown in Figure \ref{fig:1}. We note some targets were observed multiple times due to overlap that occurred in WFCAM pointings and therefore have multiple images associated with them. 

We encourage readers to examine the many WFCAM cutouts available on the ExoFOP website as well as the full images available at the WSA. The images may be particularly of interest for those planning observations of targets in C16, which is slated to begin 2017 December 7. That campaign provides the unique opportunity to observe targets simultaneously with the Kepler spacecraft and telescopes on Earth, since the spacecraft will be operating in a forward-facing direction at that time. 

\acknowledgments

We thank Watson Varricatt and Tom Kerr for assistance in planning observations and Mike Irwin for providing prompt access to processed data. 
%The WSA is part of the Wide Field Astronomy Unit hosted by the Institute for Astronomy, Royal Observatory, Edinburgh. 
%UKIRT is owned by the University of Hawaii (UH) and operated by the UH Institute for Astronomy; operations are enabled through the cooperation of the East Asian Observatory. 
When the data reported here were acquired, UKIRT was supported by NASA and operated under an agreement among the University of Hawaii, the University of Arizona, and Lockheed Martin Advanced Technology Center; operations were enabled through the cooperation of the East Asian Observatory. 
The data reported here were obtained as part of the UKIRT Service Programme.

\clearpage

%% An example figure call using \includegraphics

%% An example table using AASTeX's deluxetable. Note that since
%% only one figure OR one table is allowed this is commented out.
% \begin{deluxetable}{ccc}
% \tablecaption{K2 Targets Observed with WFCAM\label{targets}}
% \tablehead{
% \colhead{K2 Campaign} & \colhead{Targets Observed with K2} & \colhead{Targets Observed with WFCAM}
% }
% \startdata
% 4 & 15,781 & 14,722 \\
% 5 & 26,527 & 18,437 \\
% 13 & 21,367 & 528 \\
% 16 & 29,413 & 13,844 \\
% \enddata
% \tablecomments{Some targets in the K2 Campaign 18 field will also have been observed with WFCAM; however, the official target list for that campaign has not yet been released so we do not include it in the table here. In addition, we note that some targets were observed multiple times, but only the unique number of targets with WFCAM images is listed here.}
% \end{deluxetable}  

\end{document}